\documentclass[12pt]{emulateapj}
\newcommand{\jcap}{J. Cosmology Astropart. Phys., 10, 1}

\shorttitle{Merging Cluster \object{DLSCL J0916.2+2951}}
\shortauthors{Dawson et al.}

\begin{document}


\title{Discovery of a Dissociative Galaxy Cluster Merger with Large Physical Separation}


\author{William A. Dawson\altaffilmark{1}, David Wittman\altaffilmark{1}, M. James Jee\altaffilmark{1}, Perry Gee\altaffilmark{1},  John P. Hughes\altaffilmark{2},  J. Anthony Tyson\altaffilmark{1}, Samuel Schmidt\altaffilmark{1}, Paul Thorman\altaffilmark{1}, Maru\v{s}a Brada\v{c}\altaffilmark{1}, Satoshi Miyazaki\altaffilmark{3}, Brian Lemaux\altaffilmark{1}, Yousuke Utsumi\altaffilmark{3}, Vera E. Margoniner\altaffilmark{4}}

\altaffiltext{1}{University of California, Davis, Physics Department, One Shields Av., Davis, CA 95616, USA}
\altaffiltext{2}{Department of Physics and Astronomy, Rutgers University, 136 Frelinghuysen Road, Piscataway, NJ 08854-8019, USA}
\altaffiltext{3}{Department of Astronomical Science, The Graduate University for Advanced Studies, 2-21-1 Osawa, Mitaka, Tokyo 181-8588, Japan}
\altaffiltext{4}{Department of Physics and Astronomy, California State University, Sacramento, 6000 J Street Sacramento, CA 95819, USA }

\email{wadawson@ucdavis.edu}


\begin{abstract}
We present \object{DLSCL J0916.2+2951} ($z$=0.53), a newly discovered major cluster merger in which the collisional cluster gas has become dissociated from the collisionless galaxies and dark matter.
We identified the cluster using optical and weak lensing observations as part of the Deep Lens Survey. 
Our follow-up observations with {\it Keck}, {\it Subaru}, {\it Hubble Space Telescope}, and {\it Chandra} show that the cluster is a dissociative merger and constrain the dark matter self-interaction cross-section $\sigma_{\rm DM}m^{-1}_{\rm DM}\lesssim7$\,cm$^2$\,g$^{-1}$.
The system is observed at least $0.7\pm0.2$\,Gyr since first pass-through, thus providing a picture of cluster mergers 2--5 times further progressed than similar systems observed to date.
This improved temporal leverage has implications for our understanding of merging clusters and their impact on galaxy evolution.
\end{abstract}


\keywords{cosmic background radiation---dark matter---galaxies: clusters: individual (DLSCL J0916.2+2951)---gravitational lensing: weak---X-rays: galaxies: clusters}



\section{Introduction}

During a major galaxy cluster merger the collisionless dark matter (DM) and galaxies become dissociated from the cluster gas (the principal baryonic component), a process that is observable for $\sim$2\,Gyr \citep{rick01}.
These mergers have become important astrophysical probes providing a diagnostic of large scale structure (LSS) formation, empirical evidence favoring DM over modified gravity \citep{clow06}, constraints on the DM self-interaction cross-section \citep{mark04}, and constraints on the large scale matter-antimatter ratio \citep{stei08}.
They are also a suspected source of extremely energetic cosmic rays \citep{vanw10}.
A special subclass of mergers are probes of all the aforementioned science.
Such mergers: 1) are between two subclusters of comparable mass, 2) have a small impact parameter, 3) are observed during the short period when the cluster gas is significantly offset from the galaxies and DM, and 4) occur mostly transverse to the line-of-sight so that the apparent angular separation of the cluster gas from the galaxies and DM is maximized.  We term these systems \emph{dissociative mergers}. 
Given these requirements it is not surprising that only six such systems have been confirmed: \object{1E 0657-56} \citep{clow06}, \object{MACS J0025.4-1222} \citep{brad08}, \object{Abell 520} \citep{mahd07,okab08}, \object{Abell 2744} \citep{mert11}, \object{Abell 2163} \citep{okab11}, and \object{Abell 1758} \citep{rago11}. 
With such a small sample which only probes a narrow temporal window ($\sim$0.17--0.25\,Gyr post-collision\footnote{We define the time of collision to be the time of the first pericentric passage.}), it is not surprising that we still do not completely understand merging systems. 
For example, \citet{mahd07} and Jee et al. (2012, ApJ, in press) find a ``DM core'' coincident with X-rays but not with any bright cluster galaxies in the merging system \object{A520}, and \citet{mert11} find a ``ghost'' clump of gas leading a DM clump in \object{A2744}; there is no accepted explanation for either of these observations.
Simulations are needed to provide insight into these mysteries, but hydrodynamic simulations are complex and many assumptions are required \citep[see e.g.][]{spri07,pool06}: therefore observations of new and different systems will play a
crucial role in building confidence in these simulations.

\begin{figure}
\plotone{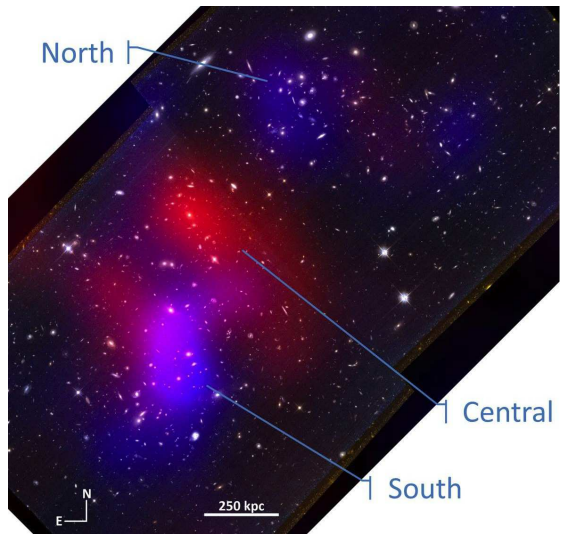}
\caption{Merging cluster DLSCL J0916.2+2951 and its three matter components. 
Overlaid on the HST color image of the galaxies is the total mass distribution (blue) based on WL analysis of the HST images and the cluster gas distribution (red) based on Chandra X-ray observations.  
The bulk of the collisional gas is located between the two collisionless galaxy and mass concentrations, indicative of a dissociative merger. 
The image is $5\arcmin \times 5\arcmin$ ($\sim 1.9\times 1.9$\,Mpc$^2$ at $z=0.53$).\label{fig1}}
\end{figure}

We have identified a new dissociative merger, \object{DLSCL J0916.2+2951}, that probes an unexplored area of merger phase-space.  
We originally detected the cluster in the Deep Lens Survey \citep[DLS;][]{DLS} via its weak lensing (WL) shear signal. 
It consists of two main subclusters  spectroscopically confirmed to be at the same redshift (0.53).
This cluster was also observed in the Sunyaev-Zel'dovich Array Survey \citep{much11} which provided evidence that the cluster gas is dissociated from the bulk of the mass and galaxies.
Follow-up optical observations with {\it Subaru} and {\it HST} enable higher resolution mass maps and follow-up X-ray observations with {\it Chandra} ACIS-I confirm that the majority of the gas is offset between the North and South subclusters, the signature of a dissociative merger (Figure \ref{fig1}).

In this letter we introduce \object{DLSCL J0916.2+2951} and summarize our survey of its three dominant components (galaxies, DM, and gas) and the cluster's astrophysical implications.
A more thorough exposition of the survey and analysis will be presented in Dawson et al. (in preparation).
Throughout this paper we assume $\Omega_{\Lambda}=0.7$, $\Omega_m=0.3$, and $H=70$\,km\,s$^{-1}$\,Mpc$^{-1}$.

\section{Optical Analysis}


We obtained spectroscopic redshifts for 20 cluster members with {\it Keck} LRIS (2007 January 16) and 634 unique spectroscopic redshifts (0\,$<$\,$z$\,$<$\,1.2)  in a $\sim15\arcmin\times 15\arcmin$ area centered on the cluster with {\it Keck} DEIMOS (2011 March 2--3), including 132 members at the cluster redshift.
We reduced the LRIS spectra using a scripted sequence of standard IRAF reduction tasks, and the DEIMOS spectra using a modified version of the DEEP2 \emph{spec2d} package \citep{davi03,gal08,lema09}.

\begin{figure*}
\epsscale{1.15}
\plottwo{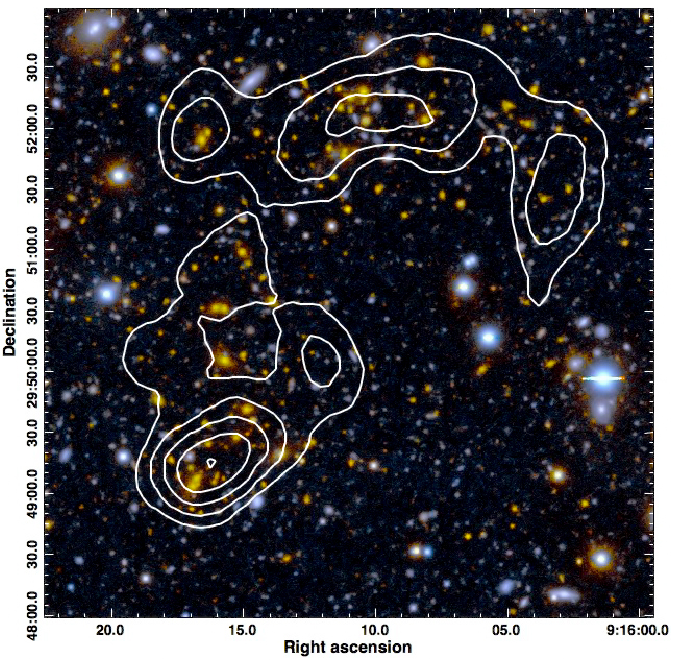}{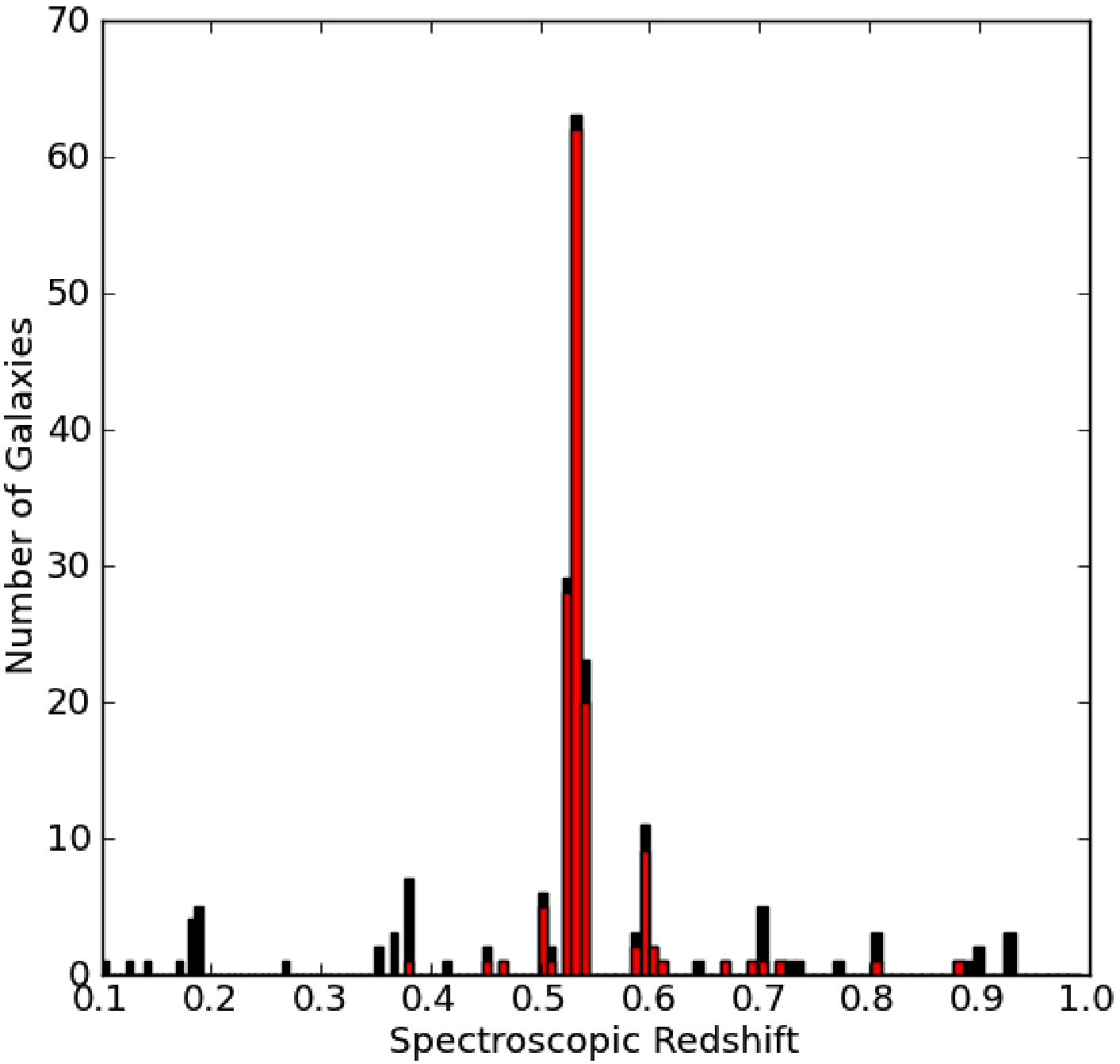}
\caption{\emph{Left:} DLS composite $BVR$ color image of DLSCL J0916.2+2951 showing  the galaxies of the two subclusters. The white contours represent the number density of galaxies with $z_{\rm phot}=0.53\pm0.1$, the cluster redshift $\pm\sigma_{z_{\rm phot}}$. The contours begin at 200 galaxies\,Mpc$^{-2}$ with increments of 50 galaxies\,Mpc$^{-2}$. 
The image field-of-view is the same as Figure \ref{fig1}.
\emph{Right:} Histogram of the 200 observed spectroscopic redshifts within the field of view of the \emph{left} figure.  The red portion is the subsample that passes the $z_{\rm phot}=0.53\pm0.1$ criteria. The galaxies at $z\sim0.6$ had equal probability of selection as the cluster members and show no sign of clustering. 
\label{fig2}}
\end{figure*}

We use our full sample of 654 spectroscopic redshifts as well as photometric redshifts to identify potential line-of-sight structures which may confuse our results.
We find no evidence for significant line-of-sight structure (Figure \ref{fig2}).

We estimate each subcluster's redshift and velocity dispersion (Table \ref{tbl1}) using the biweight-statistic and bias-corrected 68\% confidence limit \citep{beer90} applied to 100,000 bootstrap samples of each subcluster's spectroscopic redshifts.
Our redshift estimates indicate a line-of-sight velocity difference of $v_{\rm los}=670^{+270}_{-330}$ km s$^{-1}$ between the North and South subclusters, using the galaxies within a 0.5\,Mpc radius centered on the {\it HST} WL mass peaks and within a velocity range of $\pm 3000$\,km\,s$^{-1}$ of $z$=0.53 ($\sim3\times$ the expected velocity dispersion); corresponding to 38 and 35 galaxies for the North and South subcusters, respectively.
These results are robust against varying the velocity range $\pm1000$\,km\,s$^{-1}$ and using the Subaru WL or galaxy number density peaks as the apertures' centers, provided the aperture radius is $\lesssim$0.5\,Mpc: larger radii lead to significant subcluster membership confusion.
Additionally, we report the velocity dispersion mass estimates based on the scaling relation of \citet{evra08} in Table \ref{tbl1}.

We obtained $B,V,R,$ and $z'$ photometric data (12, 12, 18, and 12\,ksec, respectively) with Mosaic 1 on the KPNO 4-m {\it Mayall} telescope as part of the DLS.
To improve the accuracy of our photometric redshifts we also observed the cluster in three medium-width optical bands ($g,h,$ and $i$ from the BATC filter set), bracketing the redshifted $4000$\,\AA\, feature, using the upgraded Mosaic 1.1 imager on the KPNO {\it Mayall} with exposure times of 6\,ks per filter (2011 April 22--24).
We estimate colors using \emph{ColorPro} \citep{coe06} and redshifts using \emph{BPZ} \citep{beni00}.
We replace the standard templates with a set ``tweaked'' in a method similar to that described in \citet{ilbe06}, using spectroscopic samples from SHELS \citep{gell05} and the PRIMUS survey \citep{coil10} which overlap the DLS.
Figure \ref{fig2} shows the density isopleths of galaxies with $0.43<z_{\rm phot}<0.63$ (roughly the cluster redshift $\pm\sigma_{z_{\rm phot}}$).
This map agrees well with the distribution of spectroscopically confirmed cluster members.

\section{Weak Lensing Analysis}

To map the total mass distribution we use a version of the \citet{fisc97} method modified to include a novel tomographic signal-matched filter.
The cluster's WL shear signal, $\gamma$, depends not only on the projected surface mass over-density of the cluster, $\Delta\Sigma$, but on the relative distances of the observer, the mass, and the background galaxies:
\begin{displaymath}
\gamma=\frac{\Delta\Sigma}{\Sigma_{cr}}=\frac{\Delta\Sigma4\pi G}{c^2}\frac{D_{ls}(z_l,z_s)D_l(z_l)}{D_s(z_s)}\mathcal{H}\left(\frac{z_s}{z_l}-1\right),
\end{displaymath}
where $\Sigma_{\rm cr}$ is the critical surface density, $\mathcal{H}$ is the Heaviside step function,  and $D_l$, $D_s$, \& $D_{ls}$ are the angular diameter distances to the lens, source, and between the lens and source, respectively.
Since we do not have exact redshift measurements of the source galaxies we use each galaxy's photometric redshift probability distribution function, $p(z)$, to estimate a respective $\Sigma_{\rm cr}$ for a given lens redshift,
\begin{displaymath}
\Sigma_{cr}(z_l)\approx\langle\Sigma_{cr}(z_l)\rangle=\int\Sigma_{cr}( z_l,z_s)p(z_s)dz_s.
\end{displaymath}
In addition to weights based on shape measurement errors, we also weight the shear of each galaxy based on its $p(z)$,
\begin{displaymath}
w_{\gamma}(z_l)=\frac{1}{\int\left[\Sigma_{cr}(z_l,z_s)-\langle\Sigma_{cr}(z_l)\rangle\right]^2p(z_s)dz_s}.
\end{displaymath}
This method increases the signal-to-noise of the measurement \citep[see e.g.][]{henn05}, and more accurately accounts for the errors inherent in the photometric redshift estimates, compared to single-point estimates.
We estimate uncertainties using 100 bootstrap resamplings.

Encouraged by the DLS mass and galaxy maps we obtained higher-resolution ground and space based observations.
{\it Subaru} Suprime-Cam $i'$-band coverage of the cluster was provided by engineering-time observations of DLS Field 2 (2008 January 8).
We use the Suprime-Cam data reduction software \emph{SDFRED} \citep{yagi02,ouch04} followed by \emph{SCAMP} \& \emph{SWARP} \citep{bert02,bert06} to refine the astrometry and make the final mosaic. 
\object{DLSCL J0916.2+2951} was also observed with {\it HST} ACS/WFC using F606W and F814W filters (GO-12377, PI-W. Dawson) in a $2\times1$ pointing mosaic that covers the subclusters (Figure \ref{fig1}). 
The exposure times for F606W and F814W are 2520s and 4947s per pointing, respectively.  
We reduce this data following a method similar to that presented in \citet{jee09}.  
We measure the PSF of both datasets using the PCA method presented in \citet{jee07}.

We perform our WL analysis independently on both the Subaru and HST F814W data. The Subaru data has $0.72\arcsec$ seeing and 49 WL-quality source galaxies (i.e.\,background galaxies with measured ellipticity error  $<0.3$) per arcmin$^2$. For the mass map we use an apodizing kernel radius of $0.5\arcmin$, which can be interpreted as the effective resolution of the WL mass map.
We are able to cross-match most of the detected objects with the DLS and use the $p(z)$'s discussed in the previous section.

\begin{figure*}
\epsscale{1.15}
\plotone{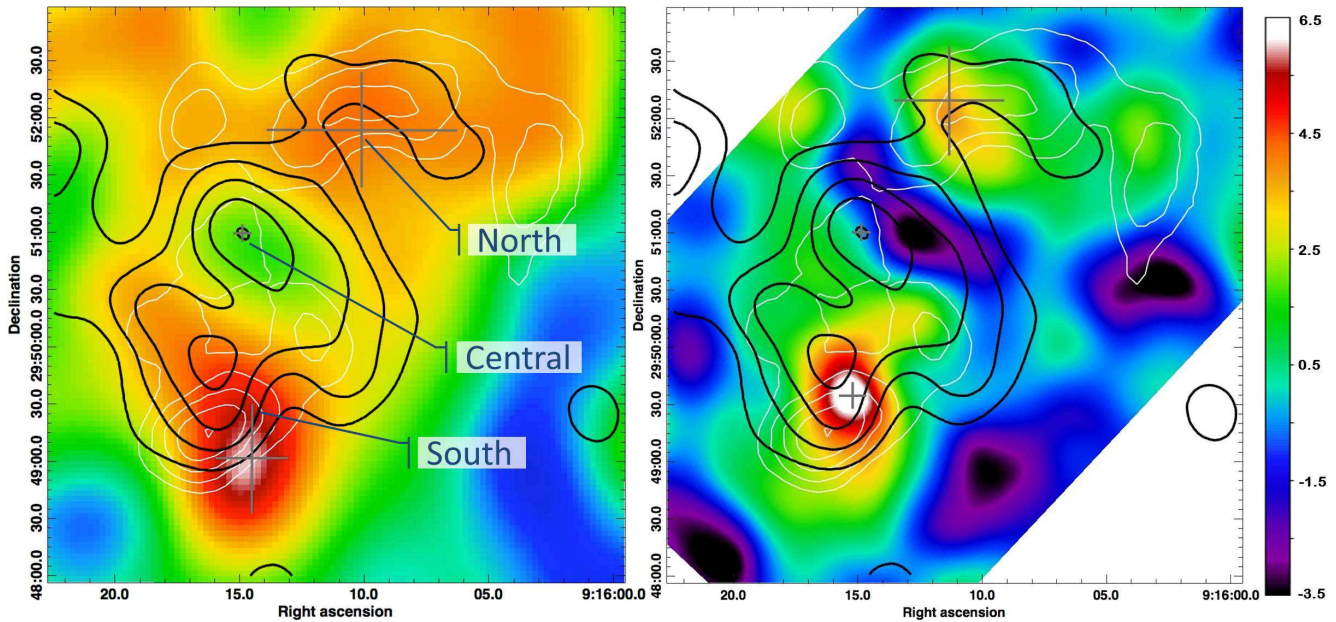}
\caption{Comparison of the Subaru $i'$-band ground-based (left) and HST space-based (right) WL mass signal-to-noise maps (color) of DLSCL J0916.2+2951 with the X-ray distribution (bold black contours) and galaxy number density (white contours, same as Figure \ref{fig2}). The peak centers and corresponding one sigma errors are denoted by the gray cross-hairs.
In both analyses there is general agreement between the location and relative magnitude of galaxies and WL yet the majority of the cluster gas is centered $\sim1.4\arcmin$ between the North and South subclusters in a local mass underdensity, providing evidence that the North and South subclusters have undergone the first pass-through of a major merger.
The scale of each map is equivalent and the image field-of-view is the same as Figures \ref{fig1} \& \ref{fig2}.
The map created from the joint Subaru/HST catalog looks nearly identical to the HST map, with only slight variations in the scale (see Table \ref{tbl1}).
\label{fig3}}
\notetoeditor{This figure should be two columns wide.}
\end{figure*}

Cross-matching is more problematic with the higher-resolution HST data, so we use a color-magnitude cut (F606W-F814W\,$<$\,0.8 and 24\,$<$\,F814W\,$<$\,28.5) to select source galaxies and exclude cluster red-sequence and bright foreground galaxies.
For the WL analysis of the HST F814W image, which has a $0.1\arcsec$ PSF and 136 WL-quality source galaxies per arcmin$^{2}$, we use an  apodizing kernel radius of $3.6\arcsec$.
We find no significant spatial correlation between source density and subcluster position, suggesting that our source galaxy population is not significantly contaminated with cluster galaxies.
We estimate the $p(z)$ of our HST source galaxy sample by assuming the photometric redshift distribution of the \citet{coe06} HUDF catalog after applying our color-magnitude cut.
Figure \ref{fig3} shows general agreement between the Subaru and HST WL mass, and galaxy density maps.

We construct a joint catalog from the HST and Subaru data, using the HST data where available and Subaru for the surrounding area.
Using a tomography-based MCMC analysis we simultaneously fit NFW halos centered on the North and South HST WL peaks, and use the \citet{gelm96} convergence test applied to eight independent chains.
In order to reduce the number of free parameters we use the \citet{duff08} empirical relation between $M_{200}$ and concentration.
We present the most likely masses for each halo along with the bias-corrected 68\% confidence limits in Table \ref{tbl1}.
We also compare the integrated projected surface mass density of the NFW halos with the measured WL aperture mass \citep{fahl94} of each subcluster and find general agreement within a radius of 0.5 Mpc of each subcluster.

\section{X-ray Analysis}
We acquired X-ray spectral-imaging of the cluster with 40ks of {\it Chandra} ACIS-I time (GO-12800854), and reduce it using \emph{CIAO} version 4.2 and \emph{CALDB} version 4.4.1.
We manually identify X-ray point sources and mask them before adaptively smoothing the diffuse emission, we present the resulting map in Figure \ref{fig3}.
We estimate source counts and their error using the \emph{dmextract} function of \emph{CIAO}.
We use an 8$\arcmin$ radius background region which encloses the subcluster regions and rests $\sim90\%$ on ACIS-I3 (on which the subclusters are observed); yet excludes subcluster, chip gaps, and point sources.
For the South and Central X-ray concentrations (120$\pm$17 and 170$\pm$19 detected 0.5-2\,keV photons, respectively) we use the \emph{Xspec} X-ray spectral fitting tool \citep{arna96} to fit a Mewe-Kaastra-Liedahl plus photoelectric absorption model \citep[fixed to the Leiden/Argentine/Bonn value;][]{kalb05} to the X-ray spectrum of each X-ray concentration.
For the North concentration there are not enough detected X-ray photons (38$\pm$12) to fit a meaningful spectrum.  
We report the results of this analysis in Table \ref{tbl1}. 

\section{Cluster Merger Scenario}
The peak of the gas distribution ($09\fh16\fm15\fs\pm5.5\fs, 29\fdg50\farcm59\farcs\pm5.0\farcs$) derived from X-rays is offset $1.4\arcmin\pm0.49$ from the North HST WL mass peak ($09\fh16\fm10\fs\pm30\fs, 29\fdg52\farcm10\farcs\pm30\farcs$), and $1.4\arcmin\pm0.14$ from the South HST WL mass peak ($09\fh16\fm15\fs\pm8.0\fs, 29\fdg49\farcm34\farcs\pm6.9\farcs$), and is located near a local minimum in the mass, suggesting that the subclusters have a small impact parameter and have experienced at least their first pass-through along a north-northwest merger direction (see Figure \ref{fig3}).
Further evidence for the merger scenario is provided by the morphology of the gas.
Simulations \citep{schi93,pool06} predict that the gas morphology elongates transverse to the merger direction after pass-through for mergers with small impact parameters.
The Central gas concentration appears to be oblate and roughly perpendicular to the axis connecting the North and South mass peaks.
This is consistent with the interpretation that these two subclusters have experienced their first pass-through and that the merger axis being roughly in the plane of the sky.

\begin{figure}
\epsscale{1.17}
\plotone{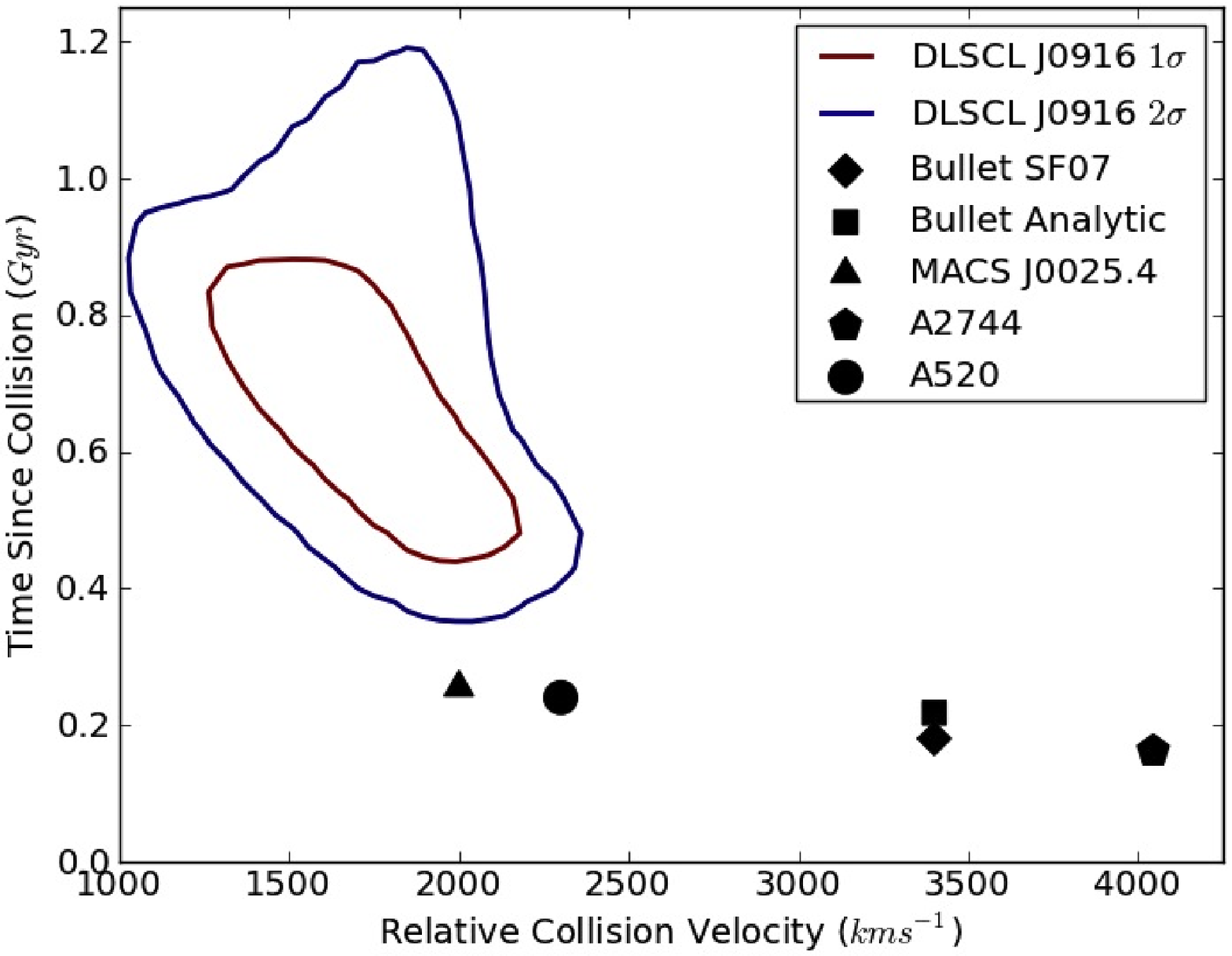}
\caption{The phase space of DLSCL J0916.2+2951 calculated from a Monte Carlo analytical treatment of the system assuming the clusters to be uniform-density spheres with $R_{200}$ radii.  The measured velocities of the other mergers \citep[with the exception of ``Bullet SF07'';][]{brad08,mark05,spri07,mert11} were input into the model to calculate their time-since-collision.  It is clear that DLSCL J0916.2+2951 probes a significantly different epoch from the existing dissociative mergers.  The close agreement between our analytical estimate, ``Bullet Analytic'', and the N-body simulation \citep{spri07}, ``Bullet SF07'', lends confidence to the results of our analytical treatment.\label{fig4}}
\end{figure}

We gain further insight into the dynamics of the system by studying the spectroscopic redshifts of the subcluster galaxies.
The observed line-of-sight velocity difference, $v_{\rm los}(t_{\rm observed})=670^{+270}_{-330}$ km s$^{-1}$, between the North and South subclusters does not explicitly determine the dynamics or orientation of the system since there is a degeneracy between the three-dimensional relative peculiar velocity, $v_{\rm relative}$, and angle of the pair with respect to the plane of the sky, $\alpha$.
In order to constrain this degeneracy we perform a Monte Carlo analysis of the system assuming the clusters to be uniform-density spheres with $R_{200}$ radii.
We draw randomly from each subcluster's mass probability distribution as determined by the NFW MCMC analysis, the redshift distribution of the spectroscopic bootstrap analysis, and the centroid distribution as determined by the bootstrap sample of the HST WL mass map.
From these parameters we calculate a conservative upper bound on the relative velocity at the time-of-collision, $v_{\rm relative}(t_{\rm collision})$, by assuming that the two subclusters freely fell from an infinite separation and zero relative velocity yet began with their current mass.
To determine a conservative lower bound we calculate $v_{\rm relative}(t_{\rm collision})$  assuming the subclusters had reached their observed separation with $v_{\rm relative}(t_{\rm observed})=0$.
Implicit in our analysis is that the merger occurs with an impact parameter of zero.
Including a moderate impact parameter of $\sim$0.1$R_{200}$ should have less than a 1\% effect on the merger velocity \citep{mast08}.
As can be seen from Figure \ref{fig4} the derived time-since-collision is relatively insensitive to moderate changes in $v_{\rm relative}(t_{\rm collision})$.
For a consistency check we compare our analytic estimate of the Bullet Cluster's time-since-collision with the numerical model results of \citet{spri07} and find agreement within $20\%$ (compare the square and diamond points of Figure \ref{fig4}).
While we attempt to be conservative in our analysis, we disregard many complexities of the merger.  
Further insight will require deeper X-ray observations and/or hydrodynamic numerical simulations.

\begin{deluxetable*}{lcccccccc}
\tablewidth{0pt}
\tabletypesize{\scriptsize}
\tablecaption{Observed subcluster and X-ray concentration properties\label{tbl1}}

\tablehead{
\colhead{Subcluster}     & \colhead{Redshift}  &
 \colhead{$\sigma_v$}    &  \colhead{$\sigma_v$ M$_{200}$}&
\colhead{WL M$_{200}$}          &
\colhead{L$_{\rm X_{\rm 0.5-2keV}}$}  & \colhead{T$_{\rm X}$}  &
\colhead{X-ray} & \colhead{Joint WL}\\
\colhead{}         &  \colhead{}  &
\colhead{(km s$^{-1}$)}    &   \colhead{($10^{14}$M$_\sun$)}  & 
\colhead{($10^{14}$M$_\sun$)}          &
\colhead{($10^{43}$erg\,s$^{-1}$)}  & \colhead{(keV)}  &
\colhead{S/N} & \colhead{S/N} 
}
\startdata
North & $0.53074^{+0.00068}_{-0.00064}$ & 
 $740^{+130}_{-190}$ &
 $3.7\pm2.3$& $1.7^{+2.0}_{-0.72}$ &
0.63 & \nodata &
 3.2 & 3.0 \\
South  & $0.53414^{+0.00065}_{-0.00064}$ & 
 $770^{+110}_{-92}$ & 
 $4.1\pm1.6$& $3.1^{+1.2}_{-0.79}$ &
2.1 & $2.7^{+1.2}_{-0.7}$ &
7.0 & 6.7\\
Central & \nodata & 
\nodata & \nodata & 
\nodata &
2.8 & $2.2^{+1.4}_{-0.6}$ &
9.1 & -3.3\tablenotemark{a}\\ 
\enddata
\tablenotetext{a}{The negative WL S/N indicates a projected surface mass local under-density.}
\end{deluxetable*}  

Using our analytic model we calculate that $v_{\rm relative}(t_{\rm collision})= 1700^{+220}_{-280}$\,km\,s$^{-1}$; given the sound speed of the gas $\sim600$\,km\,s$^{-1}$, this corresponds to a Mach number of $\sim3$.
With a projected separation of $1.0^{+0.11}_{-0.14}$\,Mpc and estimated $\alpha = 34^{+20}_{-14}$\,degrees we find a physical separation of $1.3^{+0.97}_{-0.18}$\,Mpc between the North and South subclusters.
This is the largest physical offset for a dissociative merger observed to date.
This translates into a time-since-collision ($t_{\rm observed}-t_{\rm collision}$) of $0.7^{+0.2}_{-0.1}$\,Gyr.
Given the collision velocities of the other dissociative mergers \citep{brad08,mark05,spri07,mert11} we calculate their times-since-collision using our analytic model and find that \object{DLSCL J0916.2+2951} is probing a merger stage $\gtrsim 2-5$ times further progressed than any other known dissociative merger (Figure \ref{fig4}).

\section{Dark Matter Implications}
Given the evident merger scenario we are able to use the first method of \citet{mark04} and place a rough limit on the DM self-interaction cross-section, $\sigma_{\rm DM}$.
This method compares the scattering depth of the dark matter, $\tau_{\rm DM}=\sigma_{\rm DM}m^{-1}_{\rm DM} \Sigma_{\rm DM}$, with that of the ICM gas, $\tau_{\rm ICM}\approx 1$, where $m_{\rm DM}$ is the DM particle mass and $\Sigma_{\rm DM}$ is the surface mass density of the DM particles.
$\Sigma_{\rm DM}$ is approximately the WL measured surface mass density, $\Sigma$, since $\sim80\%$ of a typical cluster's mass is DM \citep{diaf08}.
For ease of comparison with the results of \citet{mark04} and \citet{mert11} we examine the surface density averaged over the face of the subcluster within $r$=125\,kpc, which is $\Sigma\approx0.15$\,g\,cm$^{-2}$; thus we find $\sigma_{\rm DM} m_{\rm DM}^{-1} \lesssim 7$\,cm$^2$\,g$^{-1}$.  Note that we cannot apply the velocity-dependent $\sigma_{\rm DM}$ constraint methods outlined by \citet{mark04} since our analytic model assumes $\sigma_{\rm DM}$\,=\,0.

\section{Discussion}
While we use \object{DLSCL J0916.2+2951} to provide further evidence for the canonical DM model and independently constrain $\sigma_{\rm DM}m^{-1}_{\rm DM}$, we believe that its greatest value is as a probe for a new and special phase of cluster formation.  
It provides a greatly improved temporal lever-arm with which to guide numerical simulations that explore the major merger phase.  
This is potentially important given that much of our knowledge of the cluster merger process comes from numerical hydrodynamic simulations \cite[e.g.][]{pool06}, which are used to place the tightest constraints on $\sigma_{\rm DM}m^{-1}_{\rm DM}$ \citep[$<0.7$\,cm$^2$\,g$^{-1}$;][]{rand08} and bring observed merger velocities (inferred from the observed shock velocity) more in line with the expectations of $\Lambda$CDM \citep{spri07,lee10}.
Secondly, the large projected separation relative to the virial radii of the subclusters ($R_{200} \sim 1$Mpc) enables the deconvolution of the subclusters from the Central region and direct comparison of the physical properties of each.
This will provide new insight into the behavior of the cluster constituents (gas, galaxies, \& DM) during a major merger.
For example, it is well established that galaxy clusters play an important role in the evolution of their member galaxies, but it is still unclear whether cluster mergers trigger star formation \citep[e.g.][]{mill03,owen05,ferr05,hwan09}, quench it \citep{pogg04}, or have no immediate effect \citep{chun10}.
This cluster, with its estimated time-since-collision approximately equal to the lifetime of A-stars, holds promise for discriminating between slow-working processes (e.g. galaxy harassment or strangulation) and fast-acting process (e.g. ram pressure stripping). 




\acknowledgments

We thank Daniel P. Marrone, Stephen Muchovej,  John E. Carlstrom, and Tony Mroczkowski of the SZA collaboration for sharing the results of the SZA Survey, which helped motivate the detailed follow-up campaign of \object{DLSCL J0916.2+2951}. 
Support for this work was provided by NASA through Chandra Award Number GO1-12171X issued by CXO Center, which is operated by the SAO for and on behalf of NASA under contract NAS8-03060.  Support for program number GO-12377 was provided by NASA through a grant from STScI, which is operated by the Association of Universities for Research in Astronomy, Inc., under NASA contract NAS5-26555.



{\it Facilities:} \facility{CXO (ACIS-I)}, \facility{HST (ACS)}, \facility{Keck:I (LRIS)}, \facility{Keck:2 (DEIMOS)}, \facility{Mayall (MOSAIC 1 \& 1.1)}, \facility{Subaru (Suprime-Cam)}, \facility{SZA}.

\end{document}